# Formally Checking Large Data Sets in the Railways

Thierry Lecomte, Lilian Burdy[1], Michael Leuschel[2]

[1] ClearSy,
Aix en Provence, France
Thierry.lecomte@clearsy.com

[2] Formal Mind,
Dusseldorf, Germany
leuschel@cs.uni-duesseldorf.de

**Abstract.** This article presents industrial experience of validating large data sets against specification written using the B / Event-B mathematical language and the ProB model checker.

## 1 Introduction

Historically, the B Method [1] was introduced in the late 80's to design correctly safe software. Promoted and supported by RATP[1], B and Atelier B, the tool implementing it, have been successfully applied to the industry of transportation. Figure 1 depicts the worldwide implementations of the B technology for safety critical software, mainly as automatic train controllers for metros. Today, Alstom Transport Information Solutions, Siemens Transportation Systems and Technicatome-Areva are the main actors in the development of B safety-critical software. They share a product-based strategy and reuse as much as possible existing B models to develop future metros.

In the mid '90s *Event-B* [2] enlarged the scope of B to analyse, study and specify not only software, but also whole systems. *Event-B* has been influenced by the work done earlier on Action Systems [13] by the Finnish School (Action System however remained an academic project). *Event-B* is the synthesis between B and Action System. It extends the usage of B to systems that might contain software but also hardware and pieces of equipment. In that respect, one of the outcome of *Event-B* is the proved definition of systems architectures and, more generally, the proved development of, so called, "system studies" [7][8][9][10][11], which are performed before the specification and design of the software. This enlargement allows one to perform failure studies right from the beginning in a large system development. *Event-B* has been used to perform system level safety studies in the Railways [12],

---

[1] Régie Autonome des Transports Parisiens : operates bus and metro public transport in Paris

allowing to formally verify part of the whole system specification, hence contributing to improve the overall level of confidence of the railways system being built.

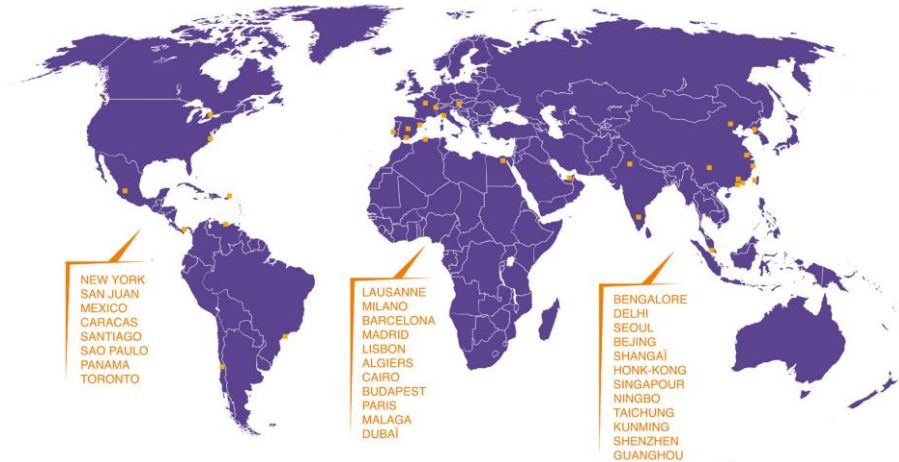

**Fig. 1.** Worldwide implementations (2012) of systems embedding software generated from B models.

However, if the verification of Event-B system specification or B software specification is quite easily reachable by semi-automated proof[2], verifying embedded data against properties[3] may turn out to be a nightmare in case of large data sets. For the Meteor metro (line 14, Paris), software and data were kept together in a B project [5]. Demonstrating data correctness regarding expected properties was really difficult as it requires to iterate over large sets of variables and constants (and their domains) and the Atelier B main theorem prover[4] is not designed for this activity [6], that requires more a model checker or constraint solver rather than a theorem prover. Later on, software and data started to be developed and validated within two different processes, in order to avoid a new compilation if the data are modified but not the software. Data validation started to be entirely human, leading to painful, error-prone, long-term activities (usually more than six months to manually check 100 000 data against 200 rules)

In this article, we present a formal approach, based on the B/Event-B mathematical language and the ProB model checker and constraint solver, designed and experimented by Alstom Transport Information Solutions for the validation of railways data.

---

[2] Automatic theorem provers usually demonstrate 90-95 % of « well written" B models, the remaining has to be demonstrated during interactive sessions with the tool.
[3] In the case of a metro, these data may represent the topology of the tracks, the position of the signals, switches, etc.
[4] That is used by both Atelier B and Rodin platform

## 2   The Genesis

Verifying railways systems covers many aspects and requires a large number of cross-verifications, performed by a wide range of actors including the designer of the system, the company in charge of its exploitation, the certification body, etc. Even if complete automation is not possible, any automatic verification is welcome as it helps to improve the overall level of confidence. Indeed a railways system is a collection of highly dependent sub-system specification and these dependencies need to be checked. They may be based on railways signalling rules (that are specific to every country or even every company in a single country), on rolling stock features (constant or variable train size or configuration) and exploitation conditions.

In France, AQL RATP laboratory initiated the development of a generic tool, OVADO[5], to verify trackside data for the metro line 1 in Paris that is being automated[6]. This tool, based on the PredicateB predicate evaluator[7], is able to parse data (XML, csv or text-based formats), load rules and verify that data complies with rules. Initially tested on line 13 configuration data, the tool has been able to check 400 definitions and 125 rules in 5 minutes. In Fig 2, we see on the left a data (called E_a_trainDynamicDeparture_minimum_speed) that associates to a train (refered to by an integer index) its minimum speed (a floating point value). It is declared as a total function, indexes and minimum speed being reachable in an excel file (A7 containing the first index and AM7 the first minimum speed). On the right, a named property is described in natural and in mathematical languages. This property may refer to data previously defined.

**Fig. 2.** Example of data definition and property

However the PredicateB tool is just a calculator able to manipulate B/Event-B mathematical language predicates: it is not able to find all possible values for any non-deterministic substitution or to find all counter-examples. Moreover the way the errors are displayed may lead to difficult analysis when the faulty predicate is complex.

---

[5] Tool for checking the B properties on railway invariants, initially developed by ClearSy
[6] A specific tool, initially developed for validating line 14 data, representing more than 300 000 lines of C++ code, was too difficult to maintain and to adapt to other lines. It was not reused for other lines.
[7] Hosted by the Rodin SVN Sourceforge service (http://rodin-b-sharp.svn.sourceforge.net)

During the DEPLOY project[8], the University of Düsseldorf and Siemens Transportation Systems have elaborated a new approach, based on the ProB model checker to dramatically reduce validation duration from about six months to some minutes [3][4]. Data is extracted from ADA source code and properties come from B models. In the case of the San Juan project, 79 files with a total of 23,000 lines of B are parsed to extract 226 properties and 147 assertions. The verification took 1017 seconds and led to the discovery of 4 false formulas. ProB was then experimented with great success on several projects: Roissy Charles de Gaule airport shuttle, Barcelona line 9, San Paulo line 4, Paris line 1 and Algiers line 1. At that occasion, ProB was slightly improved in order to deal with large scale problems and well validated in order to ease its acceptance by a certification body. However analysing false properties remains difficult. In Fig 3, a failed invariant is listed on the left (the one that is rewritten as *false*) while the counterexample is shown on the right (the values used for the data that lead to the breaking of the invariant).

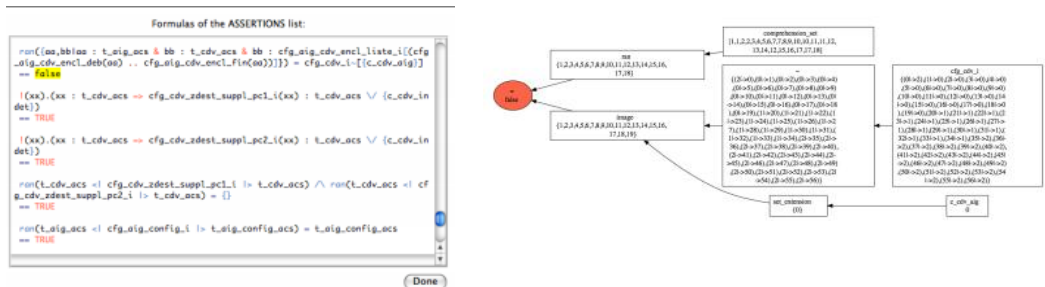

**Fig. 3.** A false property and its graphical representation

## 3 DTVT

Alstom Transport Information Solutions decided to experiment a new approach by reusing successful features of previous experiments. A new tool, DTVT, is defined and implemented. Its structure is presented in figure 4.

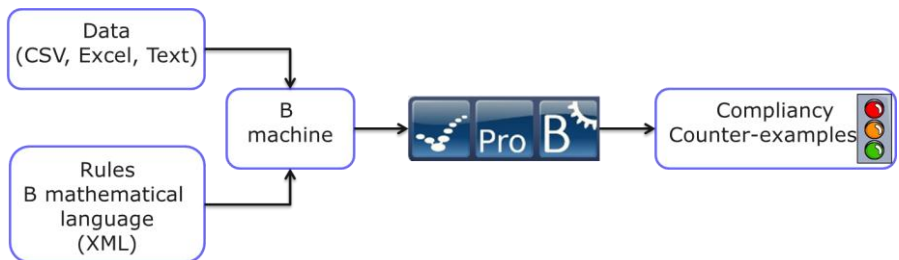

**Fig. 4.** DTVT tool structure

---
[8] http://www.deploy-project.eu/

Input data is in csv format. Data items are identified through their container file and their name. For example, *Curvatures_Cap!BeginValueCm* refers to the variable *BeginValueCm* in the file *Curvatures_Cap.xls* (see figure 5).

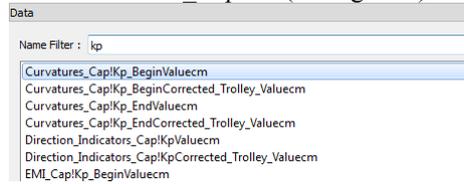

**Fig. 5.** Example of data declaration

Supported basic types are INT, BOOL and STRING. Data items are sequences of these basic types. Values are extracted from xls files (see figure 6, the positions are expressed in centimeters).

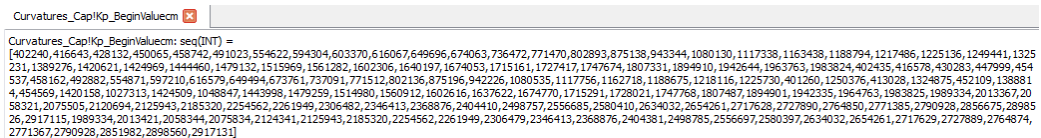

**Fig. 6.** Example of data valuation

The verification rules are expressed using the B mathematical language and structured as B operations. Instead of having to deal with too large, quantified predicates, a verification rule is decomposed in small steps that allow displaying accurate error message helping to determine the source of the error.

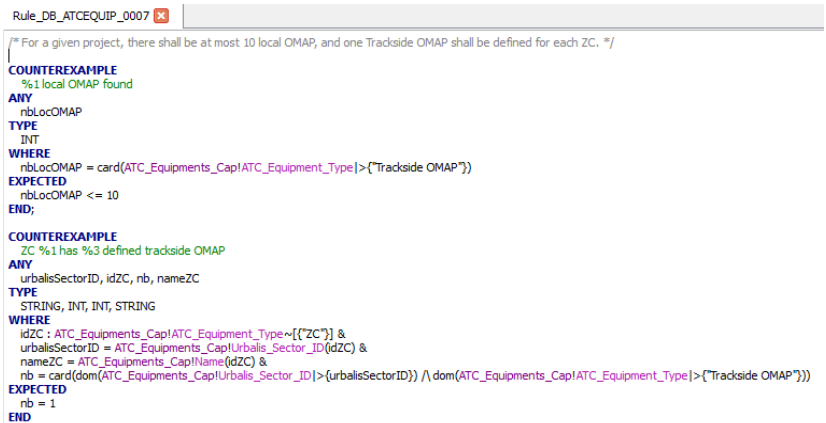

**Fig. 7.** Example of a verification rule

A rule is composed of one or several COUNTEREXAMPLEs. COUNTEREXAMPLEs are evaluated in the order they are defined. Keyword COUNTEREXAMPLE is followed by a formatted message (%1, %2, %3, etc.

represent the value of the first, second, third parameter of the following ANY substitution).

The ANY substitution allows to filter data or to calculate values. In figure 7, the first rule computes the number of couples of the sequence ATC_Equipment_Type whose second element is the string "*Trackside OMAP*".

The ANY substitution is followed by an EXPECTED field. If some values of the parameters of the ANY substitution satisfy the predicate of that substitution but don't satisfy the predicate of the EXPECTED field, the error message is displayed with its parameters instantiated. In figure 7, the error message of the second rule displays the value of *urbalisSectorID* (%1) and *nb* (%3).

ProB is the central tool for the verification. It has been modified in order to produce a file containing all counter examples detected (see figure 5) and slightly improved to better support some B keywords.

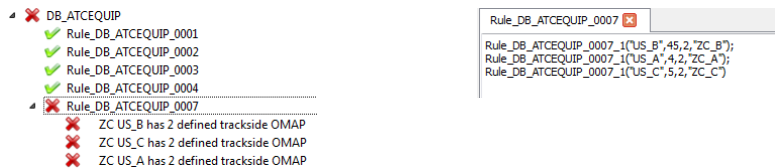

**Fig. 6.** Example of faulty verification: meaningful messages are generated for all counter examples

DTVT has been experimented with success on several ongoing developments (Mexico, Toronto, Sao Paulo, and Panama) to verify up to 50,000 Excel cells against up to 200 rules. A first round allowed defining required concepts, intermediate constructs (predicates used by several rules) and formalizing a set of generic rules that are shared by all projects. During the next rounds, specific project rules and data files were added. A complete verification is performed in about 10 minutes, including the verification report. The process is completely automatic and can be replayed without any human intervention when data values are modified.

## 5 Conclusion

Data validation appears to be of paramount importance in safety critical systems. The results obtained in this domain during the DEPLOY project have allowed to create and experiment with success on real scale projects a method for validating data against properties, based on the ProB model-checker and constraint solver.

# References

1. Abrial, J.R. (1996) , *The B-book: Assigning programs to meanings,* Cambridge University Press
2. Abrial, J.R (2005)., *Rigorous Open Development Environment for Complex Systems: event B language*
3. Leuschel, Michael and Falampin, Jérôme and Fabian, Fritz and Daniel, Plagge (2009), *Automated Property Verification for Large Scale B Models*. In: *Proceedings FM 2009*. Springer-Verlag.
4. Michael Leuschel (2012), Formal Mind, ProB, ProR and Data Validation with B, FM'2012, Industray Day.
5. Patrick Behm , Paul Benoit , Alain Faivre , Jean-marc Meynadier (1999) , *Météor: A Successful Application of B in a Large Project*
6. Milonnet C. (1999) , *B Validation Book*; Internal document ref (Matra Transport International)
7. Sabatier, D. & al (2008), *FDIR Strategy Validation with the B method*, DASIA 2008
8. Hoffmann, S. & al (2007), *The B Method for the Construction of Micro-Kernel Based Systems*, ZB 2007
9. Sabatier, D. & al (2006), *Use of the Formal B Method for a SIL3 System Landing Door Commands for line 13 of the Paris subway*, Lambda Mu 15
10. Lecomte, T. (2008), *Safe and Reliable Metro Platform Screen Doors Control/Command Systems*, FM 2008
11. Lecomte, T. & al (2007), *Formal Methods in Safety Critical Railway Systems*, SBMF 2007
12. Sabatier, D., *Formal proofs for the NYCT line 7 (Flushing) modernization project*, DEPLOY Industry Day, Fontainebleau (2012)
13. Back, R.J. & al (1991), *Stepwise Refinement of Action Systems*, Structured Programming #12 p17-30 (Springer verlag ed)